# Pauli-limited upper critical field and anisotropic depairing effect of La$_{2.82}$Sr$_{0.18}$Ni$_2$O$_7$ superconducting thin film

Ke Wang[1†], Maosen Wang[2,3†], Wei Wei[1†], Bo Hao[2,3†], Mengqin Liu[1], Qiaochao Xiang[4], Xin Zhou[1], Qiang Hou[1], Yue Sun[1], Zengwei Zhu[4], Sheng Li[1,5*], Yuefeng Nie[2,3,6*], Zhixiang Shi[1*]

1. School of Physics, Southeast University, Nanjing 211189, China
2. National Laboratory of Solid State Microstructures, Jiangsu Key Laboratory of Artificial Functional Materials, College of Engineering and Applied Sciences, Nanjing University, Nanjing 210093, People's Republic of China
3. Collaborative Innovation Center of Advanced Microstructures, Nanjing University, Nanjing 210093, People's Republic of China
4. Wuhan National High Magnetic Field Center and School of Physics, Huazhong University of Science and Technology, Wuhan 430074, China
5. Purple Mountain Laboratories, Nanjing 211111, China
6. Jiangsu Physica Science Research Center, Nanjing 210093China

* Corresponding author: sheng_li@seu.edu.cn, ynie@nju.edu.cn, zxshi@seu.edu.cn

†These authors contributed equally to this work

**Abstract:** We investigate the upper critical field and superconducting anisotropy of epitaxial La$_{2.82}$Sr$_{0.18}$Ni$_2$O$_7$ thin films, which show a sharp superconducting transition at $T_c = 31.6$ K. Near $T_c$, superconductivity exhibits thickness-limited two-dimensional characteristics. Upon cooling, the out-of-plane coherence length $\xi_c$ decreases below the sample thickness of ~ 6 nm, corresponding to a 3-unit-cell film, indicating a crossover to intrinsic three-dimensional bulk superconductivity. High-field transport measurements reveal large upper critical fields with a small anisotropy ratio $\gamma = H_{c2}^{ab}/H_{c2}^{c} \approx 1.34$, comparable to bulk Ruddlesden-Popper nickelates. At low temperatures, the in-plane (*ab*) upper critical field $H_{c2}^{ab}$ is strongly suppressed by spin-paramagnetic pair breaking and approaches the Pauli limit ($H_{c2}^{Pauli} = 58$ T), while $H_{c2}^{c}$ remains largely unaffected. This anisotropic Pauli limitation accounts for the reduced upper critical field anisotropy and supports the conclusion that superconductivity in these films is fundamentally three-dimensional bulk like. Our results highlight the essential role of spin-paramagnetic effects in shaping the high-field superconducting phase diagram of Ruddlesden-Popper nickelates.

**INTRODUCTION**

The discovery of Ruddlesden-Popper (RP) type bilayer nickelate superconductors La$_3$Ni$_2$O$_7$ has attracted considerable attention owing to their relatively high superconducting transition temperatures ($T_c$), which can exceed the liquid nitrogen range under high-pressure conditions [1-9]. Superconductivity in these nickelates shares essential features with other families of unconventional superconductors such as cuprates and iron pnictides: all possess layered crystal structures, strong electronic correlations arising from 3*d* electrons, and superconductivity that emerges in close proximity to competing magnetic and/or charge-ordered phases. These commonalities may constitute the key ingredients for achieving high $T_c$ superconductivity [10-13]. Extensive theoretical and experimental studies have been devoted to understanding the superconducting properties of this emerging system [14-21]. In RP nickelates, the application of high pressure plays a decisive role in inducing superconductivity by driving a structural phase transition from the ambient-pressure orthorhombic phase to the high-pressure tetragonal phase. This transition enhances interlayer hopping ($t$) and pushes the bonding state of the Ni $d_{z^2}$ orbital across the Fermi level, which is widely believed to be a critical factor in realizing high−$T_c$ superconductivity [17, 22-24]. Although bulk superconductivity has been established under high pressure, the extreme conditions required impose severe constraints on both fundamental investigations and potential practical applications. More recently, several independent research groups have successfully realized superconductivity at ambient pressure by growing bilayer RP nickelate thin films—such as $La_3Ni_2O_7$, $(LaPr)_3Ni_2O_7$, and $(LaSr)_3Ni_2O_7$—on $SrLaAlO_4$ substrates [25-28]. Although the transition temperatures of these films are lower than those of their high-pressure bulk counterparts, they still exceed the McMillan limit, representing a significant milestone in the exploration of unconventional superconductivity in nickelates. Detailed structural analyses reveal that compressive strain from the substrate reduces the in-plane lattice constant while elongating the out-of-plane lattice parameter [26, 28]. This strain stabilizes a crystal structure closely resembling the high-pressure I4/mmm phase of bulk samples, thereby providing a robust platform for elucidating the microscopic mechanism of high-$T_c$ superconductivity [29-36].

Unlike infinite-layer nickelate 112 thin films, which exhibit the highest transition temperature near 40 K, share the same $Ni^{1+}$ $d^9$ configuration as cuprates, and display mixed $s$- and $d$-wave gap characteristics [37-40], RP-type $La_3Ni_2O_7$ contains apical oxygens. Consequently, both $d_{x^2-y^2}$ and $d_{z^2}$ orbitals contribute to its low-energy electronic structure. Experimental observations from scanning tunneling microscopy (STM) and angle-resolved photoemission spectroscopy (ARPES) suggest that an $s^{\pm}$ pairing symmetry provides a better description of the superconducting gap behavior, although the underlying pairing mechanism remains under active debate [41-44].

A complementary approach to probing the pairing symmetry is the characterization of the upper critical field ($H_{c2}$), a crucial parameter that reflects the superconducting pairing strength and the dominant pair-breaking mechanisms. The upper critical field also provides insight into key superconducting properties, including gap magnitude, coherence length, anisotropy, and dimensionality. Low-temperature, high-magnetic-field measurements have been conducted on both thin-film and bulk nickelate superconductors to probe the anisotropy of the upper critical field. However, discrepancies remain between the reported values of the upper critical field and its anisotropy in these systems [45, 46]. A detailed analysis and reconciliation of the differences between thin films and bulk single crystals will be essential to clarify these inconsistencies and to establish a unified understanding of the Cooper pair-breaking mechanism.

It is noteworthy that the $T_c$ does not exhibit a systematic correlation with the film thickness in this thickness range [47]. However, STEM images reveal structural defects in thicker films (up to 17.6 nm in thickness), indicating degradation due to lattice relaxation [48]. Considering these factors, the properties of the films are influenced by the thickness, and it is crucial to select an appropriate film thickness to minimize strain relaxation, reduce strain inhomogeneity, and avoid structural defects, thereby ensuring a more uniform and well-defined strain state.

Limited by the low temperature and high magnetic field measurement, accurately determining $H_{c2}$ in high-$T_c$ superconductors remain challenging due to the sample inhomogeneity in nickelate thin films, which contributes to discrepancies in reported upper critical field between superconducting thin films and bulk superconductors under high pressure [5, 27, 45]. In this work, we performed resistance measurements under magnetic fields up to 58 T, enabling a detailed investigation of the upper critical field along different crystallographic directions. The precise determination of the upper critical field, combined with anisotropic magnetoresistance measurements, provides deeper insight into the nature of superconductivity and the pairing mechanism in layered nickelate superconductors.

## RESULTS
### Structure and superconductivity characterization

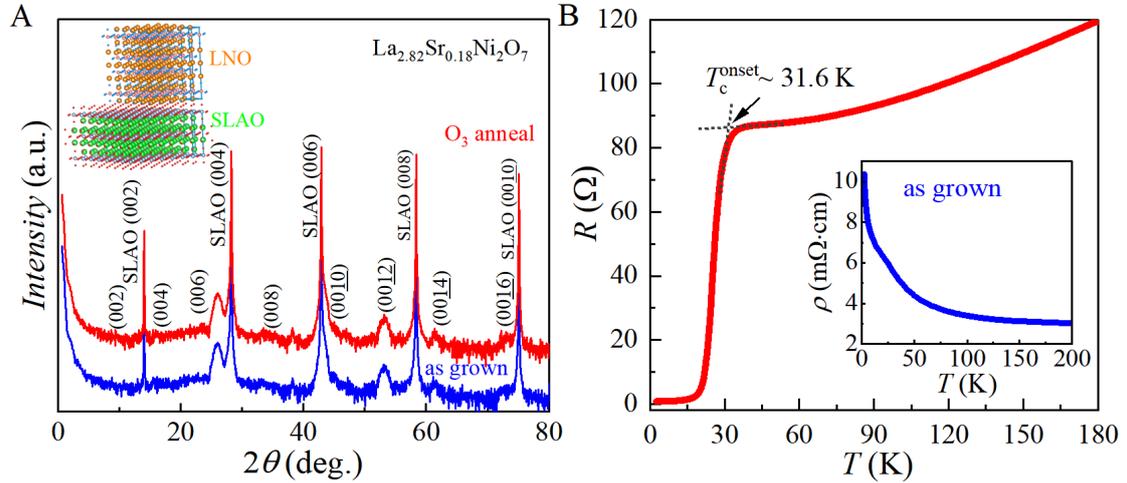

**FIG. 1: Structural and electronic properties of as-grown and ozone-annealed $La_{2.82}Sr_{0.18}Ni_2O_7$ thin film.** (A) X-ray diffraction (XRD) patterns of the as grown (blue) and annealed (red) $La_{2.82}Sr_{0.18}Ni_2O_7$ film on SLAO substrates. (B) Temperature dependent resistance with the superconducting transition ($T_c^{onset}$) at ~ 31.6 K. The inset shows the insulating behavior of the as-grown thin film.

The X-ray Diffraction (XRD) patterns of the as grown (blue) and annealed (red) $La_{2.82}Sr_{0.18}Ni_2O_7$ film on SLAO substrates have shown in Fig. 1A. Both patterns exhibit (00$l$) preferred peaks, indicating a preferred $c$ axis orientation. The crystal structure of $La_{2.82}Sr_{0.18}Ni_2O_7$ is illustrated in the inset. No discernible shifts in peak position, changes in line shapes, or variations in relative intensities are detected after the annealing process, implying the absence of significant structural modifications. The $c$-axis lattice constant is extracted as 20.78 Å through the XRD $2\theta$-$\omega$ scan, representing an elongation of approximately 2.4 % compared to the bulk value resulting from the epitaxial compressive strain. Fig. 1B shows the temperature dependence resistance of the electrical resistance. The as-grown film exhibits insulating behavior across the measured temperature range (inset of Fig. 1B). In contrast, the annealed film demonstrates metallic behavior at high temperatures, which is followed by a superconducting transition with an onset temperature ($T_c^{onset}$) of approximately 31.6 K. This $T_c^{onset}$ is determined from the intersection point of two linear fits applied to the normal and transition regions of the resistance curve. Notably, no resistance anomaly is observed near 150 K, which stands in contrast to bulk crystals [1, 3, 9, 49] . This suggests that the charge- or spin-density-wave order prevalent in bulk materials the compressive strain induced by the SLAO substrate. This strain effect is analogous to the application of high pressure to bulk crystals, where the suppression of density-wave orders also leads to the emergence of superconductivity. Above 40 K, a noticeable deviation from Fermi liquid behavior is observed, with the details of the Fermi liquid fitting shown in Fig. S1. In our nickelate superconducting thin film, reduced phase stiffness can significantly enhance superconducting fluctuations, which may influence normal-state transport well above $T_c$, leading to a deviation from Fermi liquid behavior. The combination of a sharp superconducting transition and the absence of competing electronic orders establishes this annealed thin film as an ideal platform for investigating the intrinsic superconducting properties, particularly the upper critical field.

**High-field electronic transport measurement**

The temperature dependence of the electrical resistance under magnetic fields up to 16 T, applied along the $c$ axis and $ab$ plane, have been performed in Fig. 2A and 2B respectively. The superconducting transition is markedly broadened under $H \parallel c$, whereas it remains relatively sharp under $H \parallel ab$, suggesting the presence of strong vortex thermal fluctuations. This pronounced anisotropy in transition broadening has been extensively documented in similar systems [51, 52].

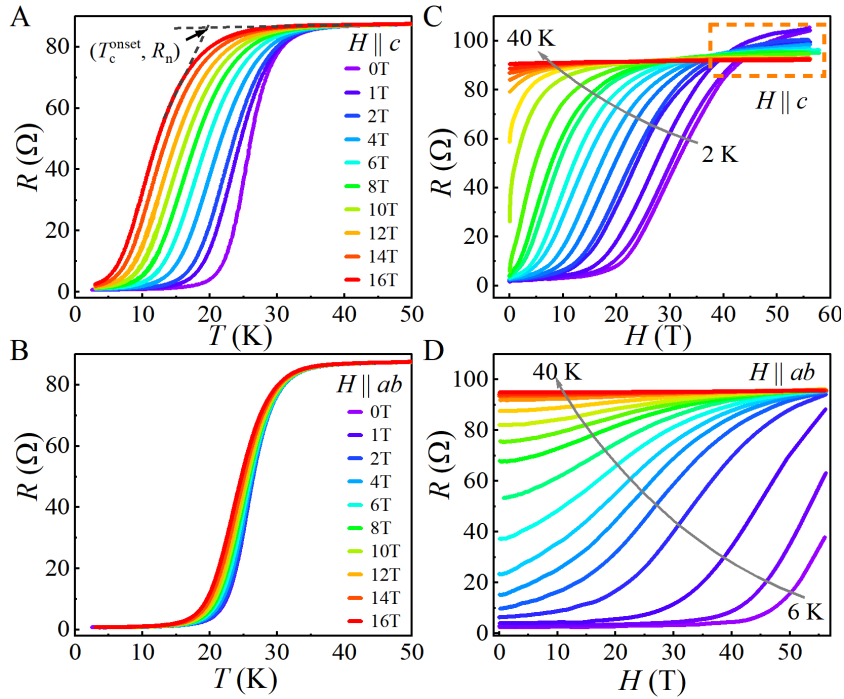

FIG. 2: **Magnetotransport properties of the La$_{2.82}$Sr$_{0.18}$Ni$_2$O$_7$ film.** (A, B) Temperature dependence of the electrical resistance under magnetic fields up to 16 T applied (A) perpendicular ($H \parallel c$) and (B) parallel ($H \parallel ab$) to the film plane. (C, D) Magnetoresistance measured up to 58 T at various temperatures for $H \parallel c$ (C) and $H \parallel ab$ (D).

Beyond the normal-state resistance upturn, we note a crossing of magnetoresistance isotherms at different temperatures (highlighted by the yellow dashed rectangle in Fig. 2C), particularly evident for $H \parallel c$. Comparable behavior has been observed in two-dimensional systems and attributed to quantum Griffiths singularity (QGS) [53, 54], linked to disorder-driven superconductor-metal transitions, and also been observed in the infinite-layer nickelate superconductor [52, 55]. Considering the inherent quenched disorder in our thin film, a similar interpretation may plausibly apply here. A less pronounced crossover is also detectable for $H \parallel ab$, likely obscured by the higher upper critical field along this orientation. However, more systematic investigations are necessary to conclusively determine the origin of this crossover behavior.

**Anisotropic upper critical field and coherent lengths**

By integrating the upper critical field ($H_{c2}$) data extracted from Fig. 2(A, C) for $H \parallel c$ and Fig. 2(B, D) for $H \parallel ab$, we construct the full temperature dependence of $H_{c2}$ over an extended temperature range. Due to significant transition broadening under applied magnetic fields, all superconducting transition temperatures are determined using the 90% $R_n$ criterion. Because of the thin-film geometry and pronounced two-dimensional behavior near $T_c$, we first applied the 2D Ginzburg-Landau (GL) model to both $H \parallel c$ and $H \parallel ab$. The resulting $H_{c2}(T)$ curves are presented in Fig. 3A, with solid lines representing fits based on the two-dimensional (2D) Ginzburg–Landau (GL) model [56]:

$$H_{c2}^c(T) = \frac{\phi_0}{2\pi\xi_{ab}^2}\left(1 - \frac{T}{T_c}\right) \quad (1)$$

$$H_{c2}^{ab}(T) = \frac{\sqrt{12}\phi_0}{2\pi\xi_{ab}(0)d_{sc}}\left(1 - \frac{T}{T_c}\right)^{\frac{1}{2}} \quad (2)$$

where $\phi_0$ is the magnetic flux quantum. The fitting yields a zero-temperature in-plane coherence length $\xi_{ab}(0) \approx 2.70$ nm and an effective superconducting thickness $d_{sc} \approx 5.11$ nm, which reasonably matches the film thickness, indicating that the entire film is superconducting, consistent with previous reports [28, 45]. The fits yield zero-temperature upper critical fields of $H_{c2}^{ab}(0\ \text{K}) = 82$ T and $H_{c2}^c(0\ \text{K}) = 45$ T. In contrast to infinite-layer nickelate superconductors, no low-temperature upturn is observed in $H_{c2}(T)$, suggesting the absence of Fulde-Ferrell-Larkin-Ovchinnikov (FFLO) state within the measured magnetic-field and temperature ranges [52, 56, 57].

Although the 2D GL model adequately describes our data, the resulting $H_{c2}$ anisotropy differs from that of bulk superconductors [18, 46]. The two-dimensional superconducting character stems from the confinement of the coherence length along the $c$ axis ($\xi_c$) within the effective superconducting thickness ($d_{sc}$). Combined with the 2D GL fitting results for $\xi_{ab}$, yielding the relation $\xi_c > d_{sc} > \xi_{ab}$, which is uncommon in layered nickelate superconductors. This discrepancy implies that $H_{c2}^{ab}$ may be overestimated by the 2D GL model, leading to the deviation from bulk behavior. To obtain more accurate coherence lengths, we analyzed $\xi_{ab}(T)$ and $\xi_c(T)$ by applying equations $H_{c2}^c = \frac{\phi_0}{2\pi\xi_{ab}^2(T)}$ and $H_{c2}^{ab} = \frac{\phi_0}{2\pi\xi_{ab}(T)\xi_c(T)}$, as shown in Fig. 3B. Solid curves represent fits based on $\xi(T) = \frac{\xi_0}{\sqrt{1-T/T_c}}$ derived from the 2D GL model. For the GL expression $\xi(T) \propto (1 - T/T_c)^{-1/2}$ is valid near $T_c$, where the condition $d_{sc} \ll \xi_c(T)$ is naturally satisfied. At lower temperatures, where this condition breaks down, a three-dimensional description is required. Accordingly, the 3D model (single-band Werthamer-Helfand-Hohenberg (WHH) theory and two-band model) is more appropriate for analyzing the low-temperature behavior. Consequently, we applied single-band WHH theory and two-band model to fit the data (Fig. 3A) along different direction.

For $H \parallel c$, the applied magnetic field is sufficient to reach $H_{c2}^c(0\text{ K})$, and $H_{c2}^c(T)$ exhibits an approximately linear temperature dependence. Such behavior is widely observed in multiband superconductors and indicates that spin-paramagnetic effects are negligible for this field orientation. Considering the multiband nature of RP nickelate superconductors, a two-band model is therefore employed to determine $H_{c2}^c$. The two-band model yields $H_{c2}^c(0\text{ K}) = 42.6$ T, in better agreement with the data and consistent with theoretically predicted multiband behavior in this system [18, 58].

For $H \parallel ab$, the limited temperature and magnetic-field range permits fitting with the single-band WHH model, yielding $H_{c2}^{ab}(0\text{ K}) = 57.1$ T. The derived coherence lengths of $\xi_{ab}(0) \approx 2.78$ nm and $\xi_c(0) \approx 2.07$ nm, are both smaller than the film thickness, supporting the scenario of bulk-like superconductivity in this thin film at low temperature. Detailed fitting parameters are provided in Table S1, and the results align well with bulk measurements. Additionally, Table S3 summarizes a comparison of the fitting results for the 2D and 3D models.

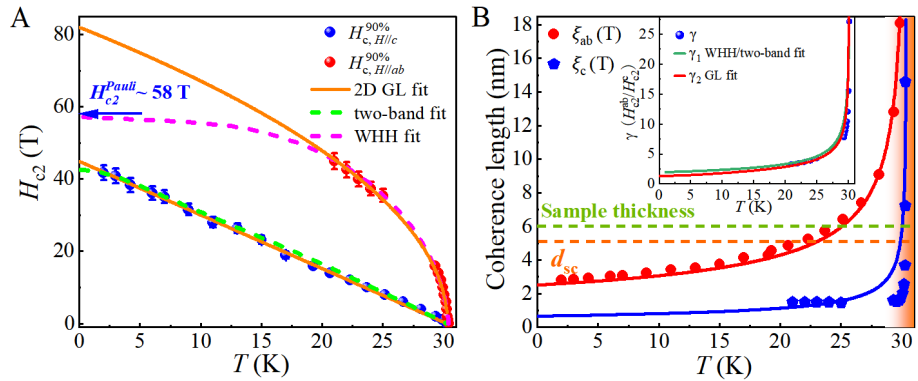

FIG. 3: **Upper critical field and coherence lengths of the La$_{2.82}$Sr$_{0.18}$Ni$_2$O$_7$ thin film.** (A) Temperature dependence of the upper critical field ($H_{c2}$) for magnetic field applied parallel ($H \parallel ab$, red) and perpendicular ($H \parallel c$, blue) to the film plane. The data are fitted using the 2D GL model, single-band WHH model, and two-band model. (B) Temperature dependence of the coherence lengths along the $c$ axis ($\xi_c(T)$) and within the $ab$ plane ($\xi_{ab}(T)$), the orange shaded region denotes the approximate 2D GL applicability range, associated with the crossover from 2D to 3D superconducting behavior. The inset shows the anisotropy ratio $\gamma$ obtained from the 3D (two-band and WHH) model ($\gamma_1$) and 2D GL model ($\gamma_2$).

Based on the fitting results in Figure 3B, the 2D GL model is only valid in the high-temperature region (near $T_c$), where the spin-paramagnetic effect is not considered, leading to a significantly higher upper critical field $H_{c2}^{ab}(0\,\text{K}) = 82$ T in the low-temperature region. For $H \parallel ab$, the strong suppression of $H_{c2}$ at low temperatures relative to the GL extrapolation provides clear evidence of Pauli-limited behavior. Therefore, we adopt the WHH model, which includes both orbital and spin-paramagnetic pair-breaking effects and yields a reduced and physically meaningful $H_{c2}^{ab}(0\,\text{K}) = 57.1$ T.

The inset of Fig. 3B displays the temperature dependence of the $H_{c2}$ anisotropy ratio $\gamma$, obtained from both 2D (GL) and 3D (two-band and WHH) model fits. Owing to the markedly different slopes between the two field orientations (Fig. 3A), $\gamma$ exhibits a divergence near $T_c$ and saturates at low temperatures, with $\gamma_{3D}(0\,\text{K}) \approx 1.34$. The strong anisotropy of $H_{c2}$ near $T_c$ is not an intrinsic property but rather a consequence of the finite film thickness ($d_{sc} \ll \xi_c$), which leads to $H_{c2}^{ab} \gg H_{c2}^{c}$ and thus an unusually large $H_{c2}$ anisotropy. As temperature decreases, $\xi_c$ becomes smaller, and once it becomes comparable to the sample thickness, superconductivity tends to exhibit intrinsic bulk-like behavior, causing $\gamma$ to decrease and approach the values observed in bulk samples [46]. At low temperatures, $\gamma$ decreases to 1.34, which we attribute to spin-paramagnetic pair-breaking effects, as discussed in detail subsequently. Furthermore, we have compiled $T_c$ and $H_{c2}$ for several different Ruddlesden-Popper nickelate materials, as summarized in Table S4. This table includes a range of materials, including bulk (both polycrystalline and single-crystal) and thin film samples, to provide a more comprehensive comparison with the experimental results obtained from our thin films.

**Angle dependent upper critical field**

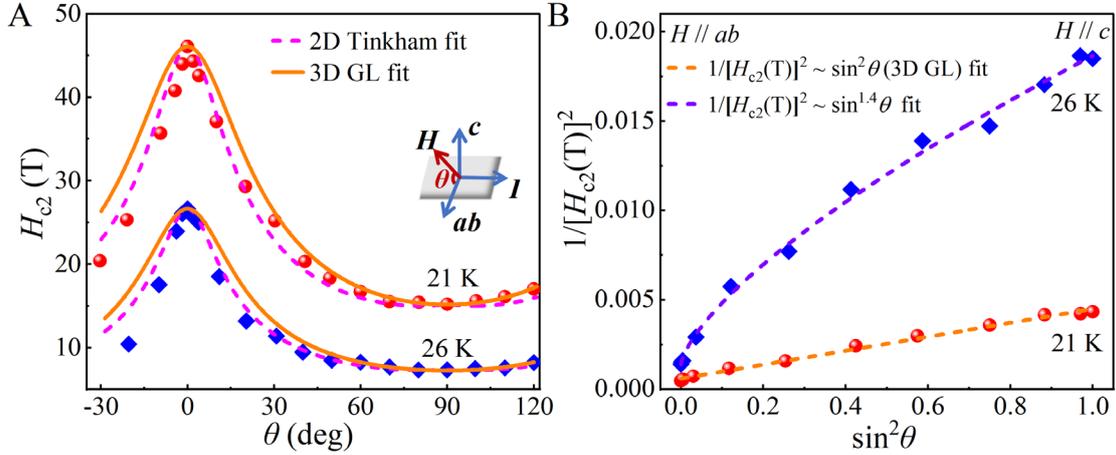

FIG. 4: **Angular-dependent upper critical field in thin film.** (A) Angular dependence of $H_{c2}(\theta)$ at 21 K and 26 K. Dashed and solid curves represent fits to the 2D Tinkham model and the 3D GL model. (B) Angular dependent $H_{c2}(\theta)$ determined from the $R$-$H$ curve at fixed temperatures, the dashed lines represent the fitting results.

To further investigate the anisotropy of the upper critical field, we performed angular-dependent $H_{c2}$ measurements at 21 K and 26 K (Fig. 4A), where $\theta = 0°$ and $\theta = 90°$ correspond to $H \parallel ab$ and $H \parallel c$, respectively. The $H_{c2}(\theta)$ data were analyzed using both the two-dimensional (2D) Tinkham model:

$$\left|\frac{H_{c2}(\theta)\sin\theta}{H_{c2}^c}\right| + \left(\frac{H_{c2}(\theta)\cos\theta}{H_{c2}^{ab}}\right)^2 = 1, \tag{3}$$

and the three-dimensional (3D) GL model:

$$H_{c2}(\theta) = H_{c2}^{ab}/\sqrt{(\gamma^2-1)\sin^2\theta + 1}. \tag{4}$$

As shown in Figs. 4A and 4B, the angular dependence of $H_{c2}$ at 26 K is well described by the 2D Tinkham model. In contrast, the data at 21 K lie between the predictions of the two models, with a linear relation between $\sin^2\theta$ and $1/H_{c2}(\theta)$ that aligns more closely with the 3D GL model. At 26 K, the relationship between $\sin^2\theta$ and $1/H_{c2}(\theta)$ deviates from linearity and is better captured by a scaling behavior between $(\sin^2\theta)^{0.7}$ and $1/H_{c2}(\theta)$. These results indicate that superconductivity in thin film exhibits 2D characteristics near $T_c$, while progressively transitioning toward intrinsic bulk like behavior at lower temperatures, consistent with the temperature evolution of the coherence length along the $c$ axis. The anisotropy of the magnetoresistance in thin films under various magnetic fields within the superconducting transition region further supports the anisotropy of the upper critical field, with detailed analyses provided in the Supplementary Materials (Figs. S4-6).

**DISCUSSION**

Our analysis indicates that the upper critical field behavior near $T_c$ is captured by the 2D GL model, while at low temperatures, the 3D two-band model and WHH model offer a more physically accurate description, supporting the scenario of bulk-like superconductivity in this thin film at low temperature.

For a superconductor, upper critical field can be limited by both orbital pair-breaking and spin-paramagnetic (Pauli limit) pair-breaking mechanisms due to copper pairs formation. The spin-paramagnetic pair-breaking originates from the Zeeman effect, where an applied magnetic field aligns electron spins, thereby breaking singlet Cooper pairs. The corresponding Pauli limit is given by

$$H_{c2}^{\text{Pauli}} = \Delta/\sqrt{2}\mu_B = 1.84 T_c \tag{5}$$

with $\mu_B$ is the Bohr magneton. This expression defines the intrinsic upper bound for the critical field in singlet-pairing superconductors, beyond which the energy required for spin polarization surpasses the condensation energy. For our $La_{1.82}Sr_{0.18}Ni_2O_7$ thin film with $T_c = 31.6$ K, the Pauli limit is $H_{c2}^{\text{Pauli}} \approx 58$ T. The measured in-plane upper critical field, $H_{c2}^{ab} = 57.1$ T, closely approaches this Pauli limit, indicating a strong spin-paramagnetic pair-breaking effect along this direction. In contrast, earlier studies on similar thin films reported $H_{c2}^{ab}$ values significantly exceeding the Pauli limit when analyzed using the 2D GL model [25, 27, 28, 45]. Our coherence length analysis confirms that the film exhibits bulk-like superconductivity at low temperatures, suggesting that the WHH model offers a more physically realistic description of $H_{c2}^{ab}(T)$, particularly in the low-temperature regime. The WHH fit yields a large Maki parameter ($\alpha_M = 21$) and a relatively small spin-orbit scattering rate ($\lambda_{so} = 0.3$), consistent with strong spin-paramagnetic pair-breaking that significantly suppresses $H_{c2}^{ab}$. Consequently, the estimated in-plane critical field at 0 K is substantially reduced from $H_{c2}^{ab}(\text{GL}) = 82\text{ T}$ to $H_{c2}^{ab}(\text{WHH}) = 57.1\text{ T}$. The unusually large Maki parameter, compared to bulk nickelate superconductors, may be partially overestimate due to the steep slope of $dH_{c2}/dT$ near $T_c$. This slope is artificially enhanced by the confinement of $\xi_c$ to the superconducting layer thickness $d_{sc}$, leading to an unphysically large orbital limiting field $H_{c2}^{\text{orb}}(0\text{ K}) = -0.693T_c(\frac{dH_{c2}}{dT})_{T=T_c}$.

For fields applied along the $c$ axis, $H_{c2}^c$ exhibits an almost linear temperature dependence across the entire measured range and remains well below the Pauli limit, indicating that orbital pair-breaking is the dominant mechanism. Given that the Fermi surface of the RP nickelates consists of both $d_{z^2}$ and $d_{x^2-y^2}$ bands, we employed a two-band model to fit the $H_{c2}^c(T)$ data (see Table S2). Assuming $\alpha_M = 0$, we extracted both intraband and interband coupling constants. The condition $\lambda_{11}\lambda_{22} > \lambda_{12}\lambda_{21}$ suggests that intraband coupling dominates the stabilization of superconductivity in this thin film, analogous to iron-based superconductors [59]. The diffusivity ratio $\eta = D_2/D_1 = 0.47$, with $D_1 = 0.665 \text{ cm}^2/\text{s}$ and $D_2 = 0.312 \text{ cm}^2/\text{s}$, indicates comparable contributions from two distinct bands. Comparing with $\alpha_M = 21$ for $H \parallel ab$, the zero Maki parameter for $H \parallel c$ means strong anisotropy of the paramagnetic pair breaking effect, which could be attributed to the spin-orientation-locked pairing mechanism proposed in our earlier work on iron-based superconductors [59]. The weak interband scattering results in the linear behavior of $H_{c2}^c(T)$ extending to low temperature. Unlike infinite-layer nickelate thin films, this material shows no upward curvature in $H_{c2}^c$ at low temperatures. This absence may be attributed to a stronger diffusivity mismatch [52, 55] or more pronounced QGS and electron localization effects, all of which may contribute to the suppression of the upper critical field $H_{c2}$ at low temperatures.

**CONCLUSION**

In summary, we have conducted magnetotransport measurements on high-quality La$_{2.82}$Sr$_{0.18}$Ni$_2$O$_7$ thin films. The film exhibits two-dimensional electronic transport behavior in the normal state, consistent with the 2D variable range hopping model. Furthermore, a crossover behavior in the magnetoresistance curves at different temperatures suggests the possible emergence of a quantum Griffiths singularity. The evolution of the coherent length along the $c$ axis indicates bulk-like superconducting behavior at low temperature. From upper critical field analysis, we obtained $H_{c2}^{ab}(0\text{ K}) = 57.1\text{ T}$ (based on the WHH model) and $H_{c2}^c(0\text{ K}) = 42.6\text{ T}$ (from the two-band model). Comparison with Pauli limit reveals distinct pair-breaking mechanisms: spin-paramagnetic effects for in-plane field and orbital pair-breaking for out-of-plane fields. This anisotropic pair-breaking effect accounts for the relatively small anisotropy ratio of $\gamma \approx 1.34$. These findings are in agreement with recent experimental and theoretical studies, demonstrate bulk-like superconducting characteristics, and provide new insights into the mechanism of unconventional superconductivity in nickelate-base thin films.

# METHODS

## Thin film growth

La$_{2.82}$Sr$_{0.18}$Ni$_2$O$_7$ thin films (thickness: 6 nm) capped with one unit cell of SrTiO$_3$ were grown on (001)-oriented SrLaAlO$_4$ (SLAO) substrates by reactive molecular-beam epitaxy (MBE) in a DCA R450 system at 720 °C under a distilled ozone atmosphere of $1 \times 10^{-8}$ Torr. Elemental fluxes were calibrated using a quartz crystal microbalance and further optimized via reflection high-energy electron diffraction (RHEED) oscillations during LaNiO$_3$ and SrTiO$_3$ co-deposition. The SLAO substrates were pre-annealed at 1000 °C in air prior to growth. Films were deposited using a shuttered, layer-by-layer sequence following established protocols.

## Post-growth ozone annealing

Post-growth, the films were cooled in ozone and ex-situ annealed at 380 °C for approximately 1 hour in a custom ozone system (AC-2025, IN USA Inc.) with an oxygen flow of ~ 60 sccm and an ozone concentration output of 10.4 %. To suppress second-phase formation, heating and cooling times were limited to 3 and 6 minutes, respectively.

## Magnetotransport measurements

Samples maintained under vacuum or ambient conditions were found to undergo continuous oxygen loss, resulting in increased resistivity and degradation of superconducting properties, and the samples were exposed to ambient conditions for less than two hours between removal from liquid nitrogen storage and loading into the measurement cryostat. Electrical transport measurements were performed with an AC excitation current of 5 μA under static magnetic fields up to 16 T, using a wet 16 T Oxford magnet and a 9 T Physical Properties Measurement System (PPMS, Quantum Design Inc.). High-field measurements were conducted at the Wuhan National High Magnetic Field Center (WHMFC) using a pulsed magnetic field with a peak of 60 T and a pulse duration of approximately 70 ms.

## Supplementary Materials

This file includes:
Supplementary Text
Figs. S1 to S6
Tables. S1 to S4
References

## Acknowledgements

We thank the Wuhan National High Magnetic Field Center for experimental support with high-field measurements. And the authors thank the Center for Fundamental and Interdisciplinary Sciences of Southeast University for the support in resistivity measurement.

**Funding:** This work was supported by the National Key R&D Program of China (Grant No. 2024YFA1408400, 2021YFA1400400), the CAS Superconducting Research Project under Grant No. [SCZX-0101], the National Natural Science Foundation of China (Grant No.12374135, 12304193, U24A2068), the Southeast University Interdisciplinary Research Program for Young Scholars, the Fundamental Research Funds for the Central Universities (Grant No. 2242025F10008), the Open Project of the National Laboratory (Grant No. WHMFC2024014), Jiangsu Funding Program for Excellent Postdoctoral Talent under Grant Number 2025ZB141.

**Author contributions:** K.W., S.L., Y.N., and Z.S. designed the experiments, conducted the primary measurements, and drafted the manuscript. M.W., B.H., and Y.N. synthesized the samples. K.W., S.L., Z.S., W.W, Q.H, and Y.S contributed to conceptual parts of the paper and provided overall guidance. M.L and Z.Z carried out part of the high-field transport measurements. Q.X and X.Z performed the low-field measurements. All authors discussed the results and commented on the manuscript.

**Competing interests**: The authors declare that they have no competing interests.

**Data and materials availability**: All data needed to evaluate the conclusions in the paper are present in the paper and/or the Supplementary Materials. Data will be made available on request.

**Keywords:** nickelate thin film, Ruddlesden-Popper phase, bulk-like superconductivity, anisotropic depairing effect, pauli limit, upper critical field

# Supplemental Materials for

# Pauli-limited upper critical field and anisotropic depairing effect of

# $La_{2.82}Sr_{0.18}Ni_2O_7$ superconducting thin film

Ke Wang[1†], Maosen Wang[2,3†], Wei Wei[1†], Bo Hao[2,3†], Mengqin Liu[1], Qiaochao Xiang[4], Xin Zhou[1], Qiang Hou[1], Yue Sun[1], Zengwei Zhu[4], Sheng Li[1,5*], Yuefeng Nie[2,3,6*], Zhixiang Shi[1*]

7. School of Physics, Southeast University, Nanjing 211189, China
8. National Laboratory of Solid State Microstructures, Jiangsu Key Laboratory of Artificial Functional Materials, College of Engineering and Applied Sciences, Nanjing University, Nanjing 210093, People's Republic of China
9. Collaborative Innovation Center of Advanced Microstructures, Nanjing University, Nanjing 210093, People's Republic of China
10. Wuhan National High Magnetic Field Center and School of Physics, Huazhong University of Science and Technology, Wuhan 430074, China
11. Purple Mountain Laboratories, Nanjing 211111, China
12. Jiangsu Physica Science Research Center, Nanjing 210093 China

Corresponding author: Sheng Li, sheng_li@seu.edu.cn, Yuefeng Nie, ynie@nju.edu.cn, Zhixiang Shi, zxshi@seu.edu.cn

This file includes:

Supplementary Text
Figs. S1 to S6
Tables. S1 to S4
References

**Fermi liquid behavior fitting**

To highlight the deviation from Fermi-liquid behavior, an enlarged view of the resistance region above $T_c$, along with a $T^2$ fitting line, as shown in Fig. S1. The resistance follows a Fermi-liquid–like $\rho(T)=\rho_0+AT^2$ dependence at higher temperature, while a clear deviation emerges below~50 K and becomes pronounced around 40 K.

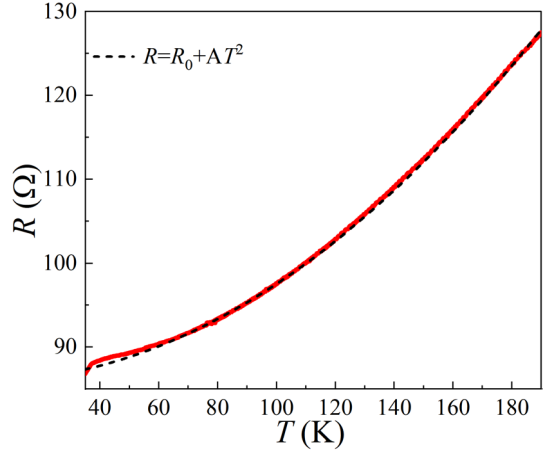

FIG. S1: Fermi Liquid Behavior Fitting. Enlarged view of the resistance region above $T_c$, along with a $T^2$ fitting line. The resistance follows a Fermi-liquid–like $\rho(T)=\rho_0+AT^2$ dependence at higher temperature.

**High-field electronic transport measurement**

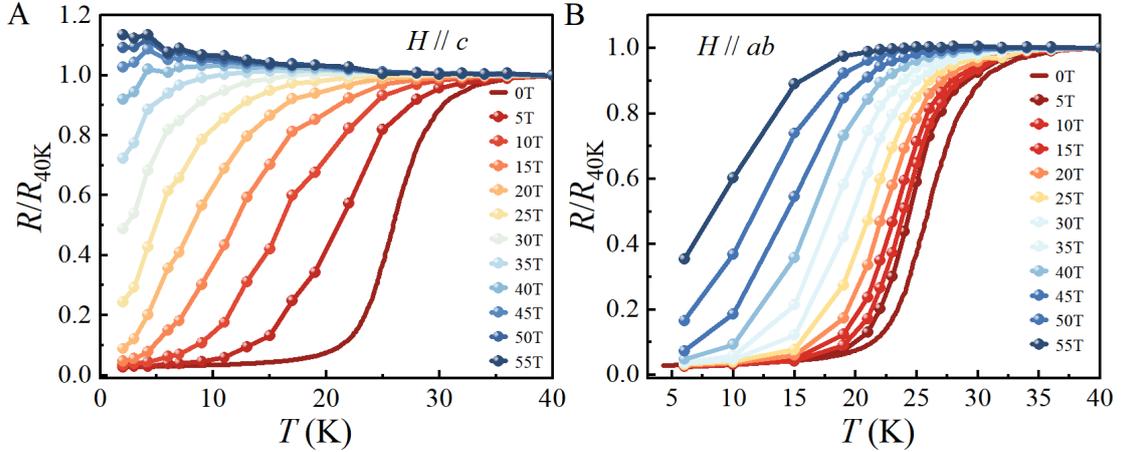

FIG. S2: Temperature dependence of resistance for $La_{2.82}Sr_{0.18}Ni_2O_7$ thin film at specific magnetic fields. Resistance versus temperature at selected magnetic fields, $H \parallel c$ (A) and $H \parallel ab$ (B). Symbols are determined from the $R$-$H$ curve at a specific magnetic field (Fig. 2C, D).

The spherical data in Fig. S2 were extracted from Figs. 2C and 2D, illustrating the temperature-dependent resistance under high magnetic fields. As shown, when the magnetic field was applied parallel to the $c$ axis ($H \parallel c$), superconductivity was completely suppressed at 55 T. In contrast, when the field was applied parallel to the $ab$ plane ($H \parallel ab$), the sample remained superconducting at 55 T. These findings are consistent with the description provided in the main text.

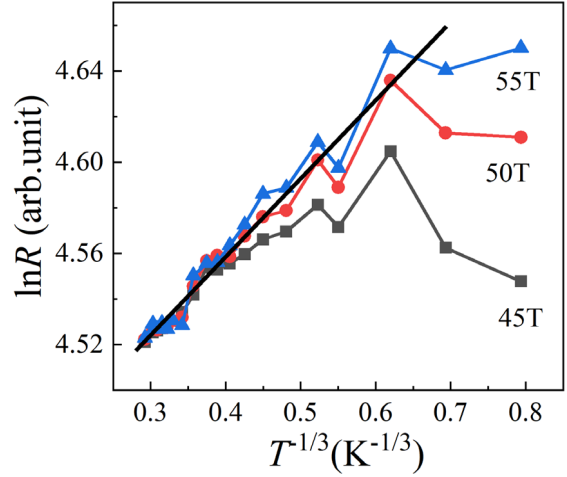

FIG. S3: The normal-state resistance under high magnetic field in Fig. S2A was fitted using the variable-range hopping (VRH) model.

The localization behavior exhibited in Fig. S2A can be effectively modeled using the variable-range hopping (VRH) model: $\rho = \rho_0 \, exp\left[\left(\frac{T_0}{T}\right)^{\frac{1}{d+1}}\right]$, where $T_0$ is a characteristic temperature and $d$ denotes the effective dimensionality. As illustrated in Fig. S3, the data are well-fitted assuming $d = 2$, indicating two-dimensional transport consistent with the thin-film geometry.

**Angular-dependent Magnetoresistance**

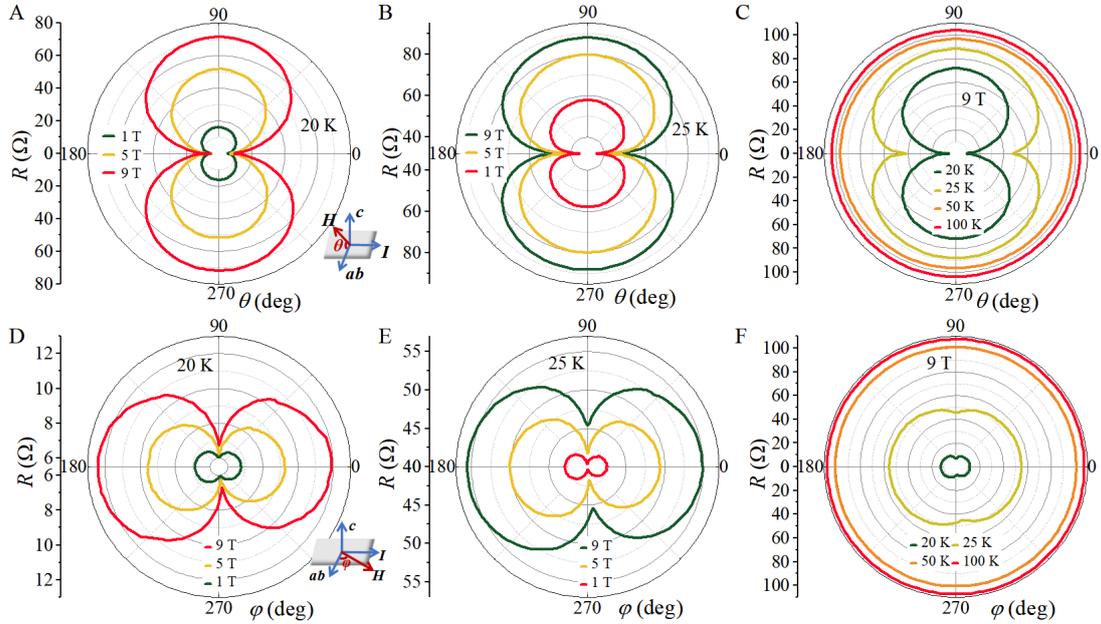

FIG. S4: Polar ($\theta$, panels A–C) and azimuthal ($\varphi$, panels D–F) angle dependence of the magnetoresistance in La$_{2.82}$Sr$_{0.18}$Ni$_2$O$_7$ thin film. (A-B) 20 K and 25 K under various magnetic fields; (C) and (F) show data at 9 T for different temperatures. (D-E) correspond to 20 K and 25 K under various fields.

In-plane magnetoresistance measurements provide a powerful probe of superconducting gap symmetry, since pair-breaking effects under high magnetic fields can reveal symmetry in the gap function $\Delta(\varphi)$. For superconductors with an anisotropic gap, the angular dependence of magnetoresistance often reflects the gap symmetry. For example, a $d$-wave superconductor typically exhibits fourfold symmetry in the in-plane magnetoresistance due to its nodal structure, whereas an $s^{\pm}$-wave superconductor with an isotropic gap would not show such symmetry unless anisotropies arise from other factors like the crystal lattice or Fermi surface. Figs. S4A and S4B show the out-of-plane angular-dependent magnetoresistance of $La_{2.82}Sr_{0.18}Ni_2O_7$ thin film at 20 K and 25 K, respectively, where $\theta = 0°$ corresponds to the magnetic field perpendicular to the $c$ axis and $\theta = 90°$ to the field parallel to the $c$ axis. To eliminate the contribution of Lorentz force, magnetic field was always perpendicular to the electric current direction. Fig. S4C presents the out-of-plane angle-dependent magnetoresistance at 9 T for different temperatures (20 K, 25 K, 50 K, 100 K). The data at 20 K and 25 K fall within the superconducting transition region, while 50 K and 100 K correspond to the normal state.

Figs. S4D and S4E display the in-plane Angular-dependent Magnetoresistance (AMR) of $La_{2.82}Sr_{0.18}Ni_2O_7$ thin film at 20 K and 25 K, respectively, where $\varphi = 0°$ corresponds to the magnetic field perpendicular to the current and $\varphi = 90°$ to the field parallel to the current. Fig. S4F shows the in-plane AMR at 9 T for different temperatures (20 K, 25 K, 50 K, 100 K). In this measurement configuration, the in-plane normal state AMR amplitude (Fig. S6F) is an order of magnitude smaller than that of the out-of-plane case (Fig. S6C), indicating that the Fermi surface can be regarded as nearly isotropic within the $ab$ plane. However, within the superconducting transition region, all AMR measurements within the superconducting transition region exhibit a pronounced twofold symmetry with a dip at the minimum. This behavior cannot be explained solely by the Lorentz force or the Fermi-surface anisotropy, which typically leads to a twofold dependence AMR but cannot be captured by such dip behavior.

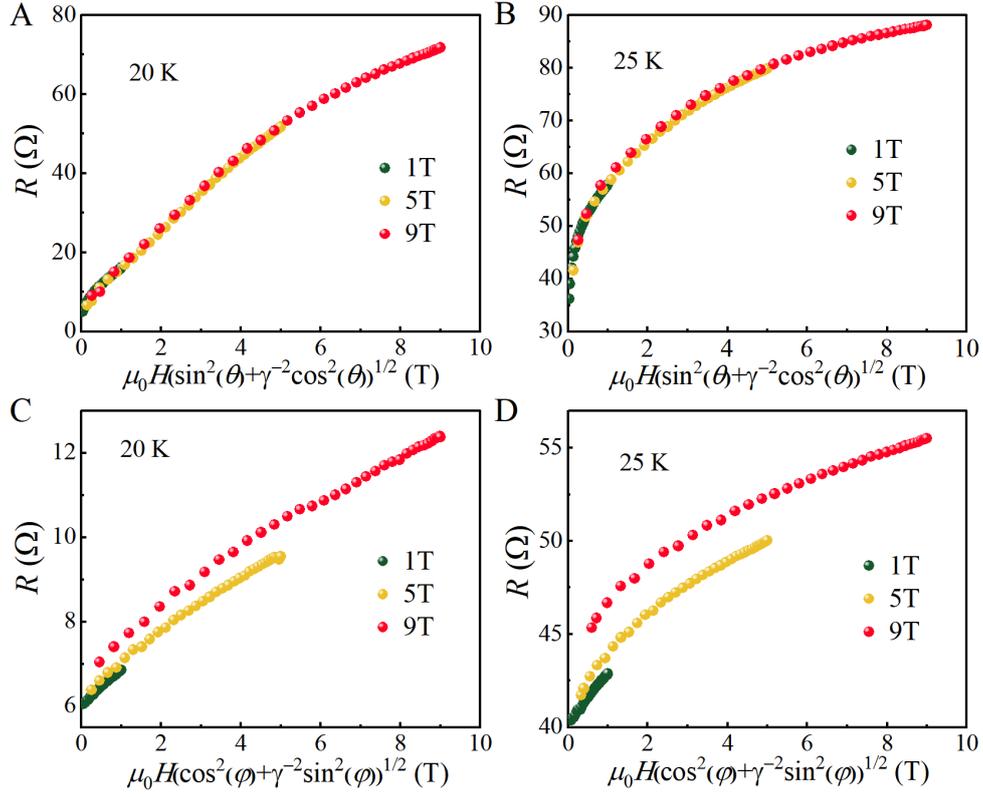

FIG. S5: Scaling behavior for Figs. S4(A-B) and (D-E). (A-B) Out-of-plane ($\theta$), Scaling behavior of the resistance vs. $\mu_0 H_s = \mu_0 H(\sin^2\theta + \gamma^{-2}\cos^2\theta)^{1/2}$, (C-D) In-plane ($\varphi$), Scaling behavior of the resistance vs. $\mu_0 H_s = \mu_0 H(\cos^2\varphi + \gamma^{-2}\sin^2\varphi)^{1/2}$.

We performed scaling analysis on the in-plane ($\theta$) and out-of-plane ($\varphi$) angular magnetoresistance data (FIG. S4). Fig. S5 shows the relation between the resistance and the scaling field $\mu_0 H_s = \mu_0 H(\sin^2\theta + \gamma^{-2}\cos^2\theta)^{1/2}$ and $\mu_0 H_s = \mu_0 H(\cos^2\varphi + \gamma^{-2}\sin^2\varphi)^{1/2}$. The results demonstrate that after adjusting the anisotropy parameter $\gamma$ of the upper critical field, the data exhibit excellent scaling behavior. The scaled out-of-plane angular data nearly collapse onto a single curve under different magnetic fields, indicating high consistency in the magnetotransport behavior along this direction. In contrast, for the in-plane angular measurements, the scaled data under various magnetic fields do not fully collapse, which can be attributed to the unavoidable influence of the Lorentz force.

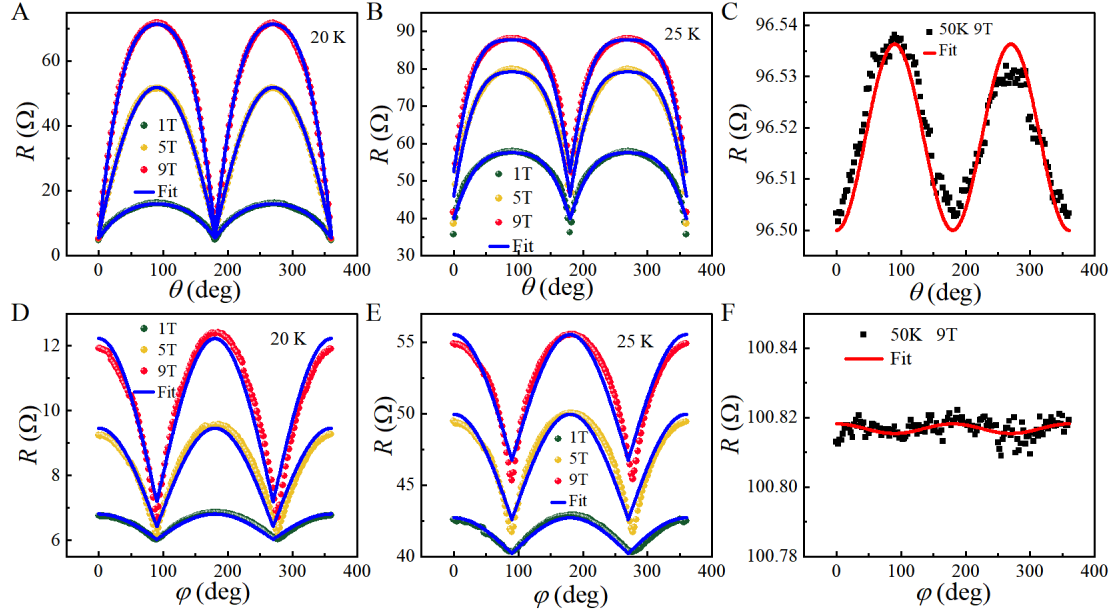

FIG S6. The angular dependence of the magnetoresistance under magnetic field up to 9 T (data sourced from Fig. S3), and the solid blue and red lines represent the fitting results. (A-C) Out-of-plane angular magnetoresistance ($\theta$), (D-F) In-plane angular magnetoresistance ($\varphi$).

We also performed fitting on the angular-dependent magnetoresistance (AMR) in the superconducting state (20 K, 25 K) and the normal state (50 K) under magnetic fields up to 9 T (data sourced from Fig. S4). Fig. S6(A-C) correspond to out-of-plane angular dependence, while Fig. S6(D-E) show the in-plane angular dependence. The out-of-plane normal-state AMR exhibits a clear two-fold symmetry, and can be well described by formula $\rho \sim \rho_0 + \Delta\rho \sin^2 \theta$, which is attributed to the twofold symmetry of Fermi pockets along $k_z$ direction. For the in-plane normal-state AMR, the observed twofold symmetry arises from the periodic variation of the Lorentz force. The normal state in-plane AMR amplitude is an order of magnitude smaller than that of the out-of-plane case, indicating that the Fermi surface can be regarded as nearly isotropic within the ab plane. The superconducting-state out-of-plane AMR data were fitted using the formula $\rho \sim \rho_0 + \Delta\rho_1 |\sin \theta| + \Delta\rho_2 \sin^2 \theta$. The twofold symmetry and the dip feature originating from the $\Delta\rho_1 |\sin \theta|$ term reflect the upper critical field anisotropy along different directions. In contrast, owing to the negligible Lorentz force contribution to AMR in the superconducting-state, the in-plane AMR data were fitted with $\rho \sim \rho_0 + \Delta\rho |\cos \varphi|$, where the observed two-fold symmetry reflects the anisotropy of the superconducting gap structure in the ab plane.

Our results revealed a twofold in plane magnetoresistance, however, due to the inevitable systematic errors in experimental measurements, the current results cannot fully rule out the contributions of Lorentz force effects and out-of-plane angular anisotropy projections to the in-plane angular magnetoresistance anisotropy signal. To accurately extract the intrinsic in-plane magnetic anisotropy information, further experimental designs are still required to effectively isolate the influence of the aforementioned factors. The absence of fourfold symmetry suggests that the nickelate superconductor is more likely to support anisotropic $s^{\pm}$-wave pairing rather than $d$-wave pairing, as previously proposed [42-44, 60-62].

**Upper critical field $H_{c2}(T)$ data fitting**

We first consider the Ginzburg-Landau (GL) model to fit the Upper critical field $H_{c2}(T)$ data, the specific fitting details have been presented in the main text. In addition, the single-band Werthamer–Helfand–Hohenberg (WHH) fitting formula for dirty limit could be used in our $H_{c2}(T)$ data analysis:

$$ln\frac{1}{t} = \sum_{\nu=-\infty}^{\infty} \{\frac{1}{|2\nu+1|} - \left[|2\nu+1| + \frac{\bar{h}}{t} + \frac{\left(\frac{\alpha_M \bar{h}}{t}\right)^2}{|2\nu+1| + \frac{\bar{h}+\lambda_{so}}{t}}\right]^{-1}\}, \quad (S1)$$

Where $\nu$ is the integer, $t = \frac{T}{T_c}$ is the reduced temperature, $\bar{h} = 2eH_{c2}\left(\frac{v_F^2\tau}{6\pi T_c}\right)$ is the reduced magnetic field, $\alpha_M$ is the Maki parameter, which represents the relative contribution of the orbital-limited and the paramagnetic-limited $H_{c2}$, and $\lambda_{so}$ is the spin-orbit scattering parameter.

For $H_{c2}(T)$, the fitting using the single-band WHH model was not successful because of the linear $H_{c2}$ behavior at low temperatures ($H \parallel c$), which is related to the multigap. Therefore, the two-band model was adopted, in the following form in the dirty limit:

$$0 = a_0[\ln t + U(h)][\ln t + U(\eta h)] + a_1[\ln t + U(h)] + a_2[\ln t + U(\eta h)], \quad (S2)$$

where $a_0 = \frac{2(\lambda_{11}\lambda_{22}-\lambda_{12}\lambda_{21})}{\lambda_0}$, $a_1 = 1 + \frac{\lambda_{11}-\lambda_{22}}{\lambda_0}$, $a_2 = 1 - \frac{\lambda_{11}-\lambda_{22}}{\frac{\lambda_0}{2}}$, $\lambda_0 = [(\lambda_{11} - \lambda_{22})^2 + 4\lambda_{12}\lambda_{21}]^{1/2}$, $t = \frac{T}{T_c}$, $h = \frac{H_{c2}^c D_1}{\frac{2\Phi_0}{T}}$, $\eta = \frac{D_2}{D_1}$, and $U(x) = \Psi\left(\frac{1}{2} + x\right) - \Psi\left(\frac{1}{2}\right)$. $\Psi(x)$ is the digamma function, $D_1$ and $D_2$ are the diffusivity of each band, $\lambda_{11}$ and $\lambda_{22}$ denote the intraband coupling constants, $\lambda_{12}$ and $\lambda_{21}$ are the interband coupling constants. Here, it is assumed that the intraband coupling dominates the $H_{c2}(T)$, and the interband coupling takes the value $\lambda_{12} = \lambda_{21}$ to reduce the number of free parameters.

Table S1. Upper critical field fitting Parameters for WHH model

| WHH fitting | $\alpha_M$ | $\lambda_{so}$ | $H_{c2}(0)$ |
|---|---|---|---|
| $H \parallel c$ | 0 | 0 | $H_{c2}^c(0) = 33.5$ T |
| $H \parallel ab$ | 21 | 0.3 | $H_{c2}^{ab}(0) = 57.1$ T |

Table S2. Upper critical field fitting Parameters for two-band model

| Two-band fitting | $D_1$ | $\eta$ | $\alpha_M$ | $\lambda_{11}$ | $\lambda_{12}(\lambda_{21})$ | $\lambda_{22}$ | $H_{c2}^c(0)$ |
|---|---|---|---|---|---|---|---|
| $H \parallel c$ | 0.665 cm²/s | 0.47 | 0 | 0.75 | 0.05 | 0.52 | 42.6 T |

**Comparison of the fitting results**

Table S3. Comparison of the fitting results for the 2D and 3D models.

| Fitting model | Fitting results | |
|---|---|---|
| 2D GL model | $H_{c2}^{ab}(0\ \text{K}) = 82\ \text{T}$<br>$H_{c2}^{c}(0\ \text{K}) = 45\ \text{T}$ | $\xi_{ab}(0) \approx 2.70\ \text{nm}$<br>$d_{sc} \approx 5.11\ \text{nm}$ |
| 3D models<br>(two-band and WHH) | $H_{c2}^{ab}(0\ \text{K}) = 57.1\ \text{T}$<br>$H_{c2}^{c}(0\ \text{K}) = 42.6\ \text{T}$ | $\xi_{ab}(0) \approx 2.78\ \text{nm}$<br>$\xi_{c}(0) \approx 2.07\ \text{nm}$ |

## Comparison of $T_c$ and $H_{c2}$

Table S4. Comparison of $T_c$ and upper critical field ($H_{c2}$) for different materials.

| | Compound | $T_c$ (K) | $H_{c2}$ | | $\gamma = H_{c2}^{ab}/H_{c2}^{c}$ | Ref |
|---|---|---|---|---|---|---|
| Film | $La_3Ni_2O_7$ | 42.4 K | $H_{c2}(0\ \text{K}) \sim 90\ \text{T}$ | | — | [25] |
| | $La_2PrNi_2O_7$ | 48.1 K | $H_{c2}(0\ \text{K}) \sim 110\ \text{T}$ | | — | [27] |
| | $La_{2.91}Sr_{0.09}Ni_2O_7$ | 38 K | $H_{c2}(0\ \text{K}) \sim 83.7\ \text{T}$ | | — | [28] |
| | $La_2PrNi_2O_7$ | 41 K | $H_{c2}^{c}(0\ \text{K})$<br>$\sim 42.8\ \text{T}$ | $H_{c2}^{ab}(0\ \text{K})$<br>$\sim 106\ \text{T}$ | 2.5 | [45] |
| | $La_{2.82}Sr_{0.18}Ni_2O_7$ | 31.6 K | $H_{c2}^{c}(0\ \text{K})$<br>$\sim 42.6\ \text{T}$ | $H_{c2}^{ab}(0\ \text{K})$<br>$\sim 57.1\ \text{T}$ | 1.34 | This work |
| Bulk | $La_3Ni_2O_7$ | 78 K | $H_{c2}(0\ \text{K}) \sim 186\ \text{T}$ | | — | [1] |
| | $La_3Ni_2O_7$ | 66 K | $H_{c2}(0\ \text{K}) \sim 97\ \text{T}$ | | — | [3] |
| | $La_4Ni_3O_{10}$ | 25 K | $H_{c2}(0\ \text{K}) \sim 44\ \text{T}$ | | — | [6] |
| | $Pr_4Ni_3O_{10}$ | 21.8 K | $H_{c2}(0\ \text{K}) \sim 27\ \text{T}$ | | — | [50] |
| | $La_4Ni_3O_{10}$ | 18.7 K | $H_{c2}^{c}(1.8\ \text{K})$<br>$\sim 24.4\ \text{T}$ | $H_{c2}^{ab}(1.8\ \text{K})$<br>$\sim 24.4\ \text{T}$ | 1 | [46] |
| | $Pr_4Ni_3O_{10}$ | 31 K | $H_{c2}^{c}(0\ \text{K})$<br>$\sim 57.3\ \text{T}$ | $H_{c2}^{ab}(0\ \text{K})$<br>$\sim 89.9\ \text{T}$ | 1.6 | [63] |